\documentclass{vgtc}                          

\ifpdf
  \pdfoutput=1\relax                   
  \usepackage{graphicx}                
  \DeclareGraphicsExtensions{.pdf,.png,.jpg,.jpeg} 
\else
  \usepackage{graphicx}                
  \DeclareGraphicsExtensions{.eps}     
\fi%

\graphicspath{{figures/}{pictures/}{images/}{./}} 
\makeatletter
\newcommand{\printfnsymbol}[1]{%
  \textsuperscript{\@fnsymbol{#1}}%
}
\makeatother
\usepackage{microtype}                 
\usepackage{enumitem}
\usepackage{xcolor}
\usepackage{hyperref}
\usepackage[compact]{titlesec}
\titlespacing{\section}{0pt}{2ex}{1ex}
\titlespacing{\subsection}{0pt}{1ex}{0ex}
\titlespacing{\subsubsection}{0pt}{0.5ex}{0ex}
\hypersetup{
  colorlinks,
  citecolor=violet,
  linkcolor=red,
  urlcolor=blue}
\PassOptionsToPackage{warn}{textcomp}  
\usepackage{textcomp}                  
\usepackage{mathptmx}                  
\usepackage{times}                     
\usepackage{cite}                      
\usepackage{tabu}                      
\usepackage{booktabs}                  

\onlineid{0}

\vgtccategory{Research}

\vgtcinsertpkg

\title{RTVis: Research Trend Visualization Toolkit}

\author{Xingyu Shen\thanks{Equal contribution}
\and Yueqian Lin\footnotemark[1]
\and Zhixian Zhang\footnotemark[1]
\and Xin Tong\thanks{Corresponding author: \href{mailto:xin.tong@dukekunshan.edu.cn}{xin.tong@dukekunshan.edu.cn}}}
\vspace{-12pt}
\affiliation{\scriptsize Duke Kunshan University}

\teaser{
  \centering
  \vspace{-12pt}	
  \includegraphics[width=0.63\linewidth]{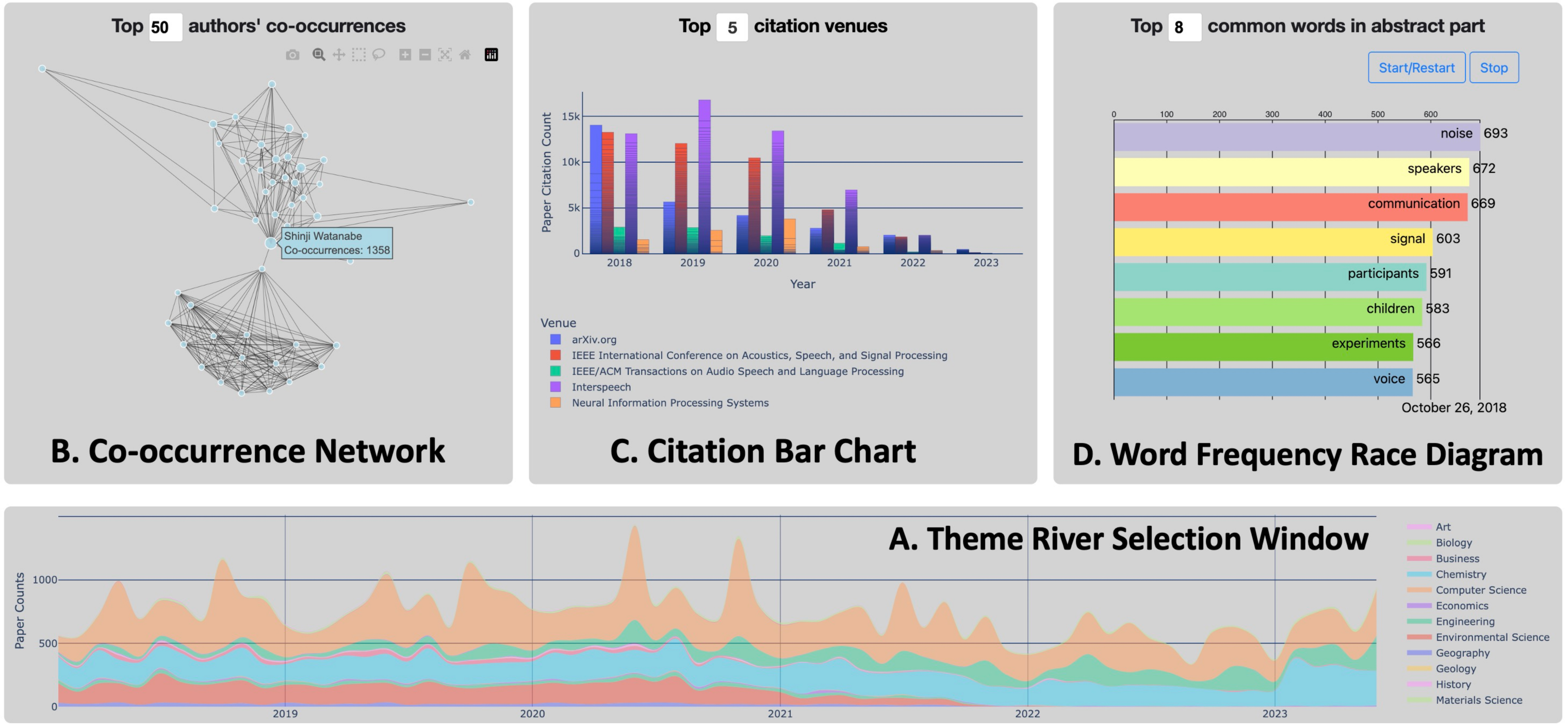}	
  \vspace{-5pt}
  \caption{RTVis: A visualization toolkit showing the research trend. \textbf{A. Theme River Selection Window} showing the number of papers in different fields over time, controlling the publication time of papers to focus on in other graphs, \textbf{B. Co-occurrence Network} showing collaborative relationships between authors, \textbf{C. Citation Bar Chart} showing the top articles with the most citations by avenue, \textbf{D. Word Frequency Race Diagram} showing the most frequently used words in abstracts.}
  \label{fig:teaser}
}
\vspace{-15pt}
\abstract{
\noindent
When researchers are about to start a new project or have just entered a new research field, choosing a proper research topic is always challenging. To help them have an overall understanding of the research trend in real-time and find out the research topic they are interested in, we developed the Research Trend Visualization toolkit (RTVis) to analyze and visualize the research paper information. RTVis consists of a field theme river, a co-occurrence network, a specialized citation bar chart, and a word frequency race diagram, showing the field change through time, cooperating relationship among authors, paper citation numbers in different venues, and the most common words in the abstract part respectively. Moreover, RTVis is open source and easy to deploy. The demo of our toolkit\footnote{\href{https://www.rtvis.design}{\color{cyan}https://rtvis.design}} and code with detailed documentation\footnote{\href{https://www.github.com/RTVis/RTVis}{\color{cyan}https://www.github.com/RTVis/RTVis}} are both available online.} 

\CCScatlist{
  \CCScatTwelve{Visualization}{Visualization techniques}{Human-centered computing}{Open-source toolkit}
}
\vspace{-5pt}

\begin{document}
\vspace{-15pt}
\firstsection{Introduction}
\maketitle
\vspace{-3pt}

\noindent
Understanding the research trend and selecting a suitable research project is paramount for researchers. However, staying abreast of rapid technological advancements and the proliferation of specialized branches within each field has made it increasingly challenging, especially for new researchers. Researchers commonly rely on reading papers as the primary approach, but gaining a comprehensive understanding of a field often requires reading dozens or even hundreds of papers, which can be a time-consuming and energy-intensive task.

To address these challenges, visual tools for analyzing research trends can provide valuable resources for researchers. While there is research on visualizing research papers \cite{Nakazawa:2015, Small:1999} that adapts the network graph to visualize paper topics and citations, these figures often tend to be complex, require time for users to understand, and can become outdated due to the lack of temporal data. Moreover, the information presented within these figures remains constrained by space limitations, providing limited insight into research trends.

Thus, we introduce RTVis, a user-friendly toolkit that offers researchers valuable insights into the development and trends within their respective fields, through various visualization tasks. We categorize the visualization tasks into four sections: exploring author collaborations, identifying highly cited papers from various venues, analyzing temporal changes in trends based on the abstracts of academic papers, and gaining a comprehensive understanding of the evolving landscape within the field. In the following section, we will explain how these functions are implemented using tools such as the \textbf{Theme River} Selection Window, Top $n$ Authors' \textbf{Co-occurrence Network}, Top $n$ Venues' Citation \textbf{Bar Chart}, and Top $n$ Word's Frequency \textbf{Race Diagram}.

\vspace{-5pt}
\section{Method}
\label{sec:method}
\vspace{-5pt}
\subsection{System Design, Interface and Interaction}
\noindent
RTVis is a system designed for facilitating academic research by supporting valid data sources and offering extensive customization options. The required data attributes for papers include title, author names, abstract, publication date, citation count, publication venue, and field of study. When choosing the visualization idioms, we followed the mental journey of researchers in selecting their research topic through a literature review: by spotting keywords to find related research papers, by reading abstracts of papers to understand the papers' main idea, by discovering authors' relationships to find more papers worth reading, and by looking at citation numbers to know the influence of papers and decide publication venues. Thus, we implemented the theme river selection window to illustrate changes in interdisciplinarity over time, the frequency race diagram to depict the abstract's hot words, the co-occurrence author network to identify clusters and key collaborators, and the citation bar chart to allow researchers to direct access to the most cited papers in the field.

We primarily built these visualizations using Plotly\footnote{\href{https://plotly.com/}{https://plotly.com/}}, along with other JavaScript components. Our goal is to achieve concise and illustrative visualizations on a single web page. To enhance understanding and interactivity, users have the ability to adjust time ranges and customize the number of items displayed in a single figure. Users can select their desired time slot on the theme river for analysis, and the toolkit will automatically retrieve published papers within that time slot and generate the corresponding visualization result. To utilize our tool, users can clone our code, prepare the dataset with the necessary headers, and instantly deploy the web app.

The following subsections describe the main graphs of the RTVis toolkit in detail:
\subsubsection{Theme River Selection Window}

We develop a theme river visualization (Figure \ref{fig:teaser}.A) following the approach described in \cite{havre2000themeriver} to depict the temporal evolution of research fields. The visualization employs a horizontal axis to represent time and a vertical axis to represent fields, presenting the changes in fields as flowing rivers. Each river corresponds to a specific field or sub-field, and the width of the river represents the number of papers in that field. Over time, the rivers representing each field shift vertically, displaying their changing positions at different time points and conveying field characteristics, such as research topics, importance, or correlations.
This visualization empowers researchers to explore research focuses and trends within various fields, providing valuable insights into the dynamics of research within specific fields.

\subsubsection{Top n Authors' Co-occurrence Network} 

The network graph (Figure \ref{fig:teaser}.B) displays academic collaborations between authors. This graph focuses specifically on the top $n$ authors based on their co-authorship frequencies, which indicates the number of authors they have collaborated with. As depicted in Figure \ref{fig:teaser}.B, the graph nodes represent authors, while the edges symbolize collaborative relationships. The size of each node is determined by its weighted degree, indicating the extent of an author's co-authorship across all papers. The graph showcases the robustness of our algorithm: first, calculating all co-occurrences and then selecting the top $n$ authors to visualize their collaborations within the selected group. This visualization enables a comprehensive analysis of the academic collaboration network, facilitating the identification of influential authors and their collaborative communities, which provides valuable insights into their interconnected relationships.

\subsubsection{Top n Venues' Citation Bar Chart}

The graph (Figure \ref{fig:teaser}.C) compares the citation counts of papers in different publication venues. Users can identify the most cited venue and paper. With a single click, they can access a direct link to the respective paper on Google Scholar. Users can choose the number of top-cited venues they wish to compare from the drop-down menu at the top. The elements in the legend are clickable, allowing users to select which venues are displayed in the chart from the selected top-cited venues. The top-cited venues are determined based on the number of citations of all papers published within the specified time range shown in the theme river. Each bar consists of multiple boxes, each corresponding to an individual paper. The box representing the paper with the highest citation count is positioned at the top of each bar. Hovering over a box reveals the title, venue, publication year, and citation count of the corresponding paper, while clicking on it opens a Google Scholar page displaying the search result for the article. Paper boxes at the bottom of the chart may appear shorter and denser. Users can select a rectangular box within the plot to view a magnified version of the smaller bars.

\subsubsection{Top n Word's Frequency Race Diagram}

The diagram (Figure \ref{fig:teaser}.D) represents each frequently occurring word in the abstract as a bar with a length proportional to its frequency. By dynamically animating their movements and heights, it effectively illustrates the fluctuations in word usage across diverse abstracts. This empowers researchers to identify pivotal topics and themes by highlighting the most frequently occurring words, providing valuable insights into the dominant interests and research directions within a specific field. Moreover, the diagram enables the comparison of word frequencies across different time periods or categories, including distinct research fields or sub-disciplines. Researchers can observe the temporal evolution of word prominence and variations across various subject areas. Additionally, we have included an input box that allows users to customize the number of bars to visualize according to their specific requirements.

\vspace{-3pt}
\section{Evaluation and Discussion}
\vspace{-3pt}
\noindent
Our toolkit thoughtfully integrates and adapts established visualization techniques to depict crucial aspects of academic papers. This integration not only ensures that the visualizations are readily accessible to users but also augments their adaptability to diverse datasets. As a result, the toolkit's applicability and potential impact are significantly broadened. For instance, beginners in a field such as speech signal processing can benefit from recognizing the trending research topics. Such an understanding is pivotal to aligning study direction with contemporary technological advancements. We have employed RTVis to distill meaningful insights from previous research to meet this aim.

Utilizing the theme river visualization, we embarked on an \textbf{insightful trend analysis} across disciplines such as computer science, medicine, and psychology. A salient observation was the incremental decline in psychology-related papers, culminating in a virtually negligible presence as of the end of 2021. Through the co-occurrence network, we were able to \textbf{identify key authors}, such as Shinji Watanabe and Brian MacWhinney, who were considered influential in their fields. Selective reading of papers related to these prominent authors and research institutions could significantly enhance efficiency, given the influential nature of their work. Furthermore, the citation bar chart furnished us with the means to \textbf{recognize prestigious venues}, including arXiv, Interspeech, and ICASSP, along with highly-cited topics like Word2vec and Conformer. This understanding resonated with insights from domain experts and elucidated where seminal work was predominantly being published. Lastly, the frequency race diagram played an indispensable role in \textbf{pinpointing the salient topics} within the research panorama. By observing the animation displaying word frequency from 2018 to 2023, it became manifest that terms associated with data science consistently prevailed. This trend suggested that utilizing machine learning methodologies to tackle speech problems was a worthwhile area of study and a mainstay in the field. 

By integrating all these visualizations, researchers could readily discern 1) the prevailing research topics, 2) the influential authors in the field, and 3) the most suitable conferences for publication. We posited that through our proposed visualizations, researchers were endowed with sufficient information to explore their selected fields.
\vspace{-5pt}
\section{Conclusion and Future work}
\label{sec:future}
\vspace{-2pt}
\noindent
We present RTVis, an open-source toolkit for understanding research trends in a specific period. RTVis allows users to visualize collaborations among authors, explore significant publication venues, analyze prevalent topics, and discover prominent research fields—all within a single web page. It supports datasets collected by users or sourced from Semantic Scholar and can be easily deployed on local machines or servers with a single run. 

While our toolkit has demonstrated promising results, there is still room for improvement. In the data processing stage, the accuracy and efficiency rely heavily on the input data quality, and data collection and cleaning are time-consuming. To enhance accuracy and efficiency, we plan to explore machine-learning approaches to automate the data-cleaning process and improve the overall reliability of the toolkit. The updating process may take longer as the system requires recalculating all components during live updates. We aim to optimize the visualization pipeline by pre-computing and caching frequently accessed calculations or visualizations to address this. This will substantially reduce the processing time required for live updates and rendering on low-computational-resource devices. Additionally, we will proactively seek user feedback through surveys and conduct user studies to identify any issues or areas for improvement. By addressing these challenges and incorporating user insights, we are confident that our toolkit will evolve to offer improved and user-friendly functionalities.
\vspace{-5pt}
\bibliographystyle{unsrt}
\bibliography{ieeevis}
\vspace{-5pt}
\newpage
\end{document}